\documentclass[12pt]{article}
\usepackage{epsfig}
\newcommand{\bea}{\begin{eqnarray}}
\newcommand{\ena}{\end{eqnarray}}

\begin{document}
\baselineskip=15pt
\begin{titlepage}
\setcounter{page}{0}

\begin{center}
\vspace*{5mm}
{\Large \bf A Tracker Solution for a Holographic Dark Energy Model}\\
\vspace{15mm}

{\large Hui Li,$^{b}$ \footnote{e-mail address: lihui@itp.ac.cn}
Zong-Kuan Guo $^{b}$ and
Yuan-Zhong Zhang$^{a,b}$}\\
\vspace{10mm}
\it $^a$CCAST (World Lab.), P.O. Box 8730, Beijing 100080, China\\
$^b$Institute of Theoretical Physics, Chinese Academy of
   Sciences, \\P.O. Box 2735, Beijing 100080, China
\end{center}

\vspace{20mm} \centerline{\large \bf Abstract} {We investigate a
kind of holographic dark energy model with the future event
horizon the IR cutoff and the equation of state $-1$. In this
model, the constraint on the equation of state automatically
specifies an interaction between matter and dark energy. With this
interaction included, an accelerating expansion is obtained as
well as the transition from deceleration to acceleration. It is
found that there exists a stable tracker solution for the
numerical parameter $d>1$, and $d$ smaller than one will not lead
to a physical solution. This model provides another possible
phenomenological framework to alleviate the cosmological
coincidence problem in the context of holographic dark energy.
Some properties of the evolution which are relevant to
cosmological parameters are also discussed.} \vspace{2mm}

\begin{flushleft}
PACS number(s): 98.80.Cq, 98.65.Dx
\end{flushleft}

\end{titlepage}

Numerous and complementary cosmological observations lend strong
support to present acceleration of our universe~\cite{Riess}. Some
negative-pressure source is needed to meet these observational
requirements. Since its energy density is unexpected small
($\rho_\Lambda\sim10^{-47}GeV^4$) in the framework of QFT, to
understand this amazing phenomenon is a great challenge to the
fundamental physics~\cite{Weinberg}. Despite a variety of
phenomenological dynamical fields (quintessence, phantom,
k-essence, etc) with suitable chosen potentials~\cite{Scalar}, the
simplest and most confusing candidate $-$ a cosmological constant
(CC) $\Lambda$ $-$ is still attractive. As a matter of fact, this
is a long-term topic and several proposals have been discussed in
the past twenty years, such as the screening of CC due to Hawking
radiation or the existence of an infrared fixed point in effective
theories of gravitation~\cite{Mottola}, the relaxation of CC based
on the coincidence limit of the graviton propagator growing in
time~\cite{Ford} and the screening mechanism due to virtual
gravitons~\cite{Wood}. Nevertheless, part of these mechanisms have
some ability to be compatible with subsequent observations of type
Ia Supernovae which indicate present cosmic acceleration, although
most of them were primarily focused on inflationary cosmology. The
theoretical enchantment of $\Lambda$ and recent remarkable
observational evidence on the equation of state (EOS) $w_{DE}$
around $-1$~\cite{Weinberg, Alam} makes theorists try to reconcile
a dynamical dark energy (DE) with a true cosmological constant $-$
hereafter we mean $w_{DE}=-1$. The congruence of these two aspects
has been stimulated these days. In Ref.~\cite{Bran} a kind of
back-reaction-induced cancellation mechanism is trying to
understand the seemingly varying CC. With the Hubble parameter $H$
the renormalization scale, a picture of the running CC has been
developed in Ref.~\cite{Shapiro}. And we note that some
phenomenological approaches to study the decaying vacuum
cosmology~\cite{Freese, PWang} have also emerged.

 On the other hand, the holographic principle~\cite{Hooft} has been inspiring great
 endeavors on attack of this fine-tuning quantity $\rho_\Lambda$. In a seminal paper
of Cohen~\cite{Cohen}, there is a suggestion that in QFT a short
distance cutoff is related to a long distance cutoff due to the
limit set by formation of a blackhole, and this ever neglected IR
limitation to QFT will correspond to an energy scale of
$10^{-2.5}eV$ if the IR cutoff is our present Hubble scale
$H_0^{-1}\approx10^{28}cm$. According to Refs.~\cite{Cohen}-\cite
{Horvat}, this choice of IR cutoff will alleviate the cosmological
constant problem~\cite{Weinberg}. In line with this suggestion,
Hsu~\cite{Hsu} and Li~\cite{Li} argued that this energy density
could be viewed as the holographic DE density satisfying
$\rho_{DE}=3 d^2M_P^2L^{-2}$, where $M_P$ is the reduced Planck
mass. Li also demonstrated that only identifying $L$ with the
radius of the future event horizon $R_e$, we can get the EOS
$w_{DE}<-1/3$ and an accelerating universe. Generally speaking,
the universe was in the matter dominated era until recently. As a
consequence, today's particle horizon and Hubble horizon are
roughly of the same order~\cite{Pea}. Then for Li's model to share
the merit of Hubble scale cutoff on reproducing the correct
magnitude of cosmological constant, the following question
emerges: why is the future event horizon $R_e$ comparable to the
particle horizon and the Hubble horizon at present? It may be
regarded as a holographic variation of cosmological coincidence
problem (CCP), which concerns about the mysterious approximate
equality of matter and DE density today~\cite{Stein}.

   Those suggestions on holographic DE have been extensively
discussed in the past few years~\cite{Horvat, Huang, Zimdahl,
Wang}. It was found that after considering the interaction between
dark matter and DE, the choice of $H^{-1}$ as the IR cutoff of the
holographic DE may be compatible with the desired
$w_{DE}=-1$~\cite{Horvat}. Nevertheless, with that interaction
included, the scaling behavior of the CC which is in favor of the
CCP may also be obtained. After loosing the constraint on
$w_{DE}=-1$, Ref.~\cite{Zimdahl} has also realized the
acceleration and a scaling solution. And there the transition from
deceleration to acceleration requires a varying coefficient $d$.
Lately, a model which uses $R_e$ instead of $H^{-1}$ as IR cutoff
and leaves $w_{DE}$ undetermined has also recovered the
acceleration transition and moreover an EOS transition from
$w_{DE}>-1$ to $w_{DE}<-1$ phantom regimes~\cite{Wang} ( see
also~\cite{quintom}).

It can be seen that, studies up to now have shown the scaling
solution in the Hubble horizon criterion of holographic DE, no
matter whether the EOS is fixed to be $-1$. However, the future
event horizon cutoff which is used to scale the holographic DE
clearly leads to a finale of $\Omega_{DE}=1$~\cite{Li}, where
$\Omega_{DE}$ is the DE fraction of the total energy density. This
is unable to address the CCP. In this Letter, we study a
particular holographic DE model~\cite{Bauer} and exhibit the
theoretical possibility to drive an $R_e$ version of holographic
DE to understand the CCP. The basic ingredients are as follows:
the holographic DE scales as $R_e^{-2}$, but different from
previous models it is assumed to possess a constant EOS
$w_{\Lambda}=-1$, i.e., we here deal with a varying but ``true"
CC.


The holographic DE density is

\begin{equation}
\rho_{\Lambda}
\equiv 3 d^2 M^2_p
R^{-2}_e,
\label{rhoL}
\end{equation}
here we keep $d$ as a free positive dimensionless parameter and
$R_e$ is the proper size of the future event horizon,
\begin{equation}
\label{Re} R_e(t) \equiv a(t) \int^{\infty}_{t} {d\tilde{t} \over
\tilde{a}(\tilde{t})} = a \int^{\infty}_{a} {d\tilde{a} \over
\tilde{H} \tilde{a}^2}\,,
\end{equation}
where $a$ is the scale factor of the universe. For a spatially
flat, isotropic and homogeneous universe with an ordinary matter
and dark energy, the Friedmann equation can be written as
\begin{equation}
\label{Friedmann} \Omega_{\Lambda} + \Omega_m = 1, \quad \Omega_m
\equiv {\rho_m \over \rho_{cr}} \quad \hbox{and} \quad
\Omega_{\Lambda}\equiv {\rho_{\Lambda} \over \rho_{cr}}\,,
\end{equation}
where $\rho_m$($\rho_{\Lambda}$) is the energy density of matter
(dark energy) and the critical density $\rho_{cr} = 3 M^2_p H^2$.
Because of the conservation of the energy-momentum tensor, the
evolution of the energy of matter and DE are governed by
\begin{eqnarray}
\label{rhomdot} \dot\rho_{\Lambda}&=&Q \nonumber
\\ \dot\rho_m+3 H \rho_m &=&-Q
\end{eqnarray}
respectively. Here we have used the requirement
\begin{equation}
\label{EOS} w_{\Lambda}= -1
\end{equation}
and $Q$ represents the undetermined interaction between matter and
DE.

By definition Eq.~(\ref{rhoL}), we have
\begin{equation}
\label{Re2} R^{2}_e\equiv\frac{3 d^2
M_p^2}{\rho_{\Lambda}}=\frac{d^2}{\Omega_{\Lambda}H^2}\,.
\end{equation}
According to the definition of the future event horizon
(\ref{Re}), a straightforward calculation can give
\begin{equation}
\label{Redot} \dot R_e=H R_e-1.
\end{equation}
Hereafter the superscript dot denotes the derivative with respect
to the cosmic time $t$. Then the rate of change of both energy
components may be expressed as
\begin{equation}
\label{rhoLdot} \dot\rho_{\Lambda}=6 M_P^2 H^3
\Omega_{\Lambda}\left(\frac{\sqrt{\Omega_{\Lambda}}}{d}-1\right) ,
\end{equation}
and
\begin{equation} \label{rhomdot} \dot\rho_m = -6M_P^2 H^3
\Omega_{\Lambda}\left(\frac{\sqrt{\Omega_{\Lambda}}}{d}-1\right)-9
M_P^2 H^3(1-\Omega_{\Lambda})
\end{equation} where
Eq.~(\ref{Friedmann}) has been recalled. It's clear that the
energy density of matter and DE can not be conservative
respectively in our model. There is energy transfer between those
two energy components and the coupling term $Q$ is just of the
form $H \rho_{\Lambda}$ multiplied by a variable coefficient. In
the present framework of our model, the truly independent
continuity equation is Eq.~(\ref{rhomdot}) and it may be employed
to produce the evolution equation for $\Omega_{\Lambda}$. By means
of Eq.~(\ref{Friedmann}), Eq.~(\ref{rhomdot}) can be cast into
\begin{equation}
 \Omega_{\Lambda}'=-3\Omega_{\Lambda}^2+
\frac{2}{d}\Omega_{\Lambda}\sqrt{\Omega_{\Lambda}}
+\Omega_{\Lambda}, \label{primeL}
\end{equation}
where the prime denotes the derivative with respect to $x\equiv
\ln a$ and then $\dot\Omega_{\Lambda}=\Omega_{\Lambda}'H$. To
study the scaling behavior of the cosmological evolution, it's
convenient to introduce an auxiliary quantity~\cite{Guo}
\begin{equation}
r\equiv
\frac{\rho_m}{\rho_{\Lambda}}=\frac{1-\Omega_{\Lambda}}{\Omega_\Lambda}.
\label{ratio}
\end{equation}
The rate $\dot r$ of the energy density ratio $r$ of matter and
dark energy can be written as
\begin{equation}
\dot r=\left(\frac{\rho_m}{\rho_{\Lambda}}\right)^{.}
 =-\frac{\dot\Omega_{\Lambda}}{\Omega_{\Lambda}^2}.
\label{rdot1}
\end{equation}
Then Eq.~(\ref{primeL}) becomes
\begin{equation}
\label{scaling} \dot r=\frac{H (3 d \Omega_\Lambda- 2
\sqrt{\Omega_\Lambda}-d)}{d \Omega_\Lambda}=0.
\end{equation}
where $\dot r=0$ gives the possible cosmological scaling behavior
$\sqrt{\Omega_\Lambda^{+}}=(1+\sqrt{1+3d^2})/(3d)$. When the
parameter $d$ is greater than 1, the positive root
$\sqrt{\Omega_{\Lambda}^{+}}$ is smaller than 1 and then a
meaningful scaling solution and vice versa. Moreover, the larger
$d$ is, the smaller value such a scaling solution can take. When
the DE component fraction is smaller than the physical scaling
solution $\Omega_\Lambda^{+}$, then $\dot r<0$ and
$\Omega_\Lambda'>0$ and as a result the DE fraction will
monotonically increase up to the maximum $\Omega_\Lambda^{+}$. In
practice, this is most relevant to our universe since the DE
component is expected to play a more and more role in the
evolution with the flow of cosmological time. As a consequence,
the observational data of present DE component fraction
$\Omega_{\Lambda}^0$ will give an observational upper bound of
$d\leq2\sqrt{\Omega_\Lambda^{0}}/(3\Omega_\Lambda^{0}-1)$ by way
of the $\Omega_{\Lambda}^0\leq\Omega_\Lambda^{+}$. For example, if
we choose $\Omega_{\Lambda}^0=0.7$, then the parameter $d$ should
not be larger than $1.5$.  People might worry about whether the
parameter $d$ can take such values greater than $1$ since the
original bound $L^3\rho_{\Lambda}\leq LM_P^2$ proposed by
Cohen~\cite{Cohen} will be violated. Careful analysis suggests
this may not be the case. The model we suggest is only a
phenomenological framework and it's unclear whether it's
appropriate to tightly constrain the value of $d$ by means of the
analogue to the blackhole physics. As a matter of fact, the
possibility of $d>1$ has been seriously dealt with and a modest
value of $d$ larger than one could be favored in the
literature~\cite{dgtone}. When $d$ equals to $1$, the positive
root $\Omega_{\Lambda}^{+}=1$ and the cosmic expansion approaches
a de Sitter phase asymptotically. How about the value $d<1$?
Unlike the original holographic dark energy model~\cite{Li, Huang}
with $R_e$ the IR cutoff where the universe approaches a phantom
phase for $d<1$, numerical simulations indicates that there exists
no consistent physical solution. In fact, $d<1$ always makes
$\Omega_\Lambda'$ positive and $\dot r<0$, even though the
increasing DE fraction has reached $1$ with matter component
vanishing. This would then ensue from an unacceptable negative
matter density. We should note that, some previous fits to the
observational SN Ia data~\cite{Huang, Wang} in the context of
holographic dark energy were on the basis different from our model
and the best fits which suggest a free parameter $d$ smaller than
$1$ may be irrelevant to present model.

In order to study the stability of the critical point of
Eq.~(\ref{primeL}) which corresponds to the scaling solution
$\Omega_{\Lambda}^{+}$, substituting a linear perturbation
$\Omega_{\Lambda} \to \Omega_{\Lambda}^{+}+\delta$ about the
critical point into the Eq.~(\ref{primeL}), to first-order in the
perturbation, gives
\begin{equation}
\delta'=\left(-6\Omega_\Lambda^{+}+\frac{3}{d}\sqrt{\Omega_\Lambda^{+}}+1\right)\delta.
\label{perturbation}
\end{equation}
It is easy to check that the scaling solution $\Omega_\Lambda^{+}$
is always the late-time stable attractor solution.

Now let's turn to more details of this model which are relevant to
some other observational quantities. The transition of
deceleration to acceleration happened when
\begin{equation}
\label{acceleration} \frac{\ddot a}{a} = - \frac{1} {6 M^2_P}
(\rho_{\Lambda} + 3 p_{\Lambda} + \rho_m) = 0.
\end{equation}
Using Eqs.~(\ref{Friedmann}) and (\ref{EOS}) as well as the above
formula we can obtain that the transition emerges at
$\Omega_\Lambda^T=1/3$, which is irrelevant to the parameter $d$.
This result coincides with that of the standard LCDM scenario
since both models have the same EOS of minus one.

The deceleration parameter $q$ is
\begin{equation}
q\equiv
-\frac{\ddot{a}
a}{\dot
a^2}=-\frac{
H'}{H}-1. \label{q}
\end{equation}
Through Eqs.~(\ref{Re}) and (\ref{Re2}), we have
\begin{equation}
\frac{d}{\sqrt{\Omega_{\Lambda}}H a}=\int^\infty_a
\frac{d\tilde{a}}{\tilde{H} \tilde{a}^2}\label{int}=\int^\infty_x
\frac{d\tilde{x}}{\tilde{a}}
\end{equation}
and taking derivative with respect to $\tilde{x}$ in both sides of
the above equation we get
\begin{equation}
\label{primeH}
\frac{H'}{H}=\frac{\sqrt{\Omega_{\Lambda}}}{d}-\frac{\Omega_{\Lambda}'}{2
\Omega_{\Lambda}}-1.
\end{equation}
Using the information extracted from Eq.~(\ref{primeL}), the
deceleration parameter $q$ may be determined by virtue of
Eqs.~(\ref{q}) and (\ref{primeH}). If exhibit the evolution of
this model in redshift, we may note that $1+z=1/a$ (by
conventional we have chosen present scale factor $a_0=1$) and then
there is a relation $x=-\ln(1+z)$.

\begin{figure}
\begin{center}
\includegraphics[width=11cm]{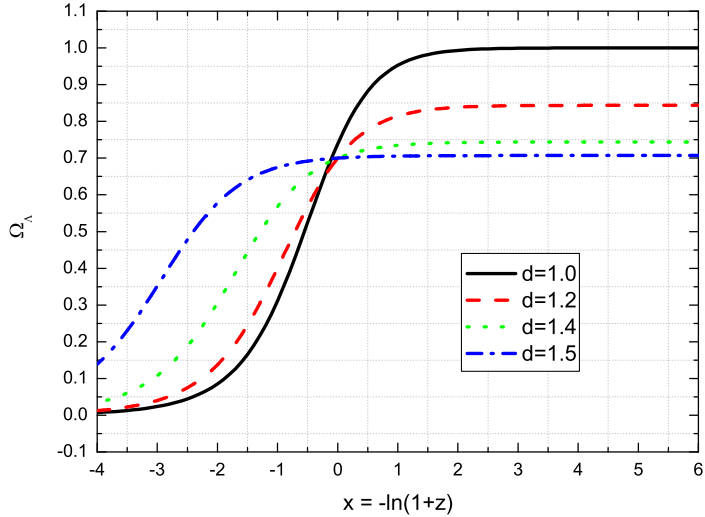}
\caption{Evolution of the DE for different values of the constant
$d$.} \label{fig:omegax}
\end{center}
\end{figure}

After specifying the value of the parameter $d$, the behavior of
the DE evolution may be obtained through Eq.~(\ref{primeL}). The
dependence of the evolution of DE with respect to the constant $d$
is shown in Fig.~\ref{fig:omegax}. We see that for different
values of the parameter $d>1$, the evolution of the universe will
approach different tracker solutions. The solution
$\Omega_{\Lambda}^{+}$ is illustrated as the plateau of a
particular curve in Fig.~\ref{fig:omegax}. What's more, the larger
$d$ is, the more gently $\Omega_\Lambda$ climb up to an end value
and the less such a value. For $d=1$, the evolution will approach
a de Sitter universe and the energy component of matter will be
infinitely diluted. Once again, we should note that, in our model
$d<1$ would yield an unphysical solution and then is not allowed.

\begin{figure}
\begin{center}
\includegraphics[width=11cm]{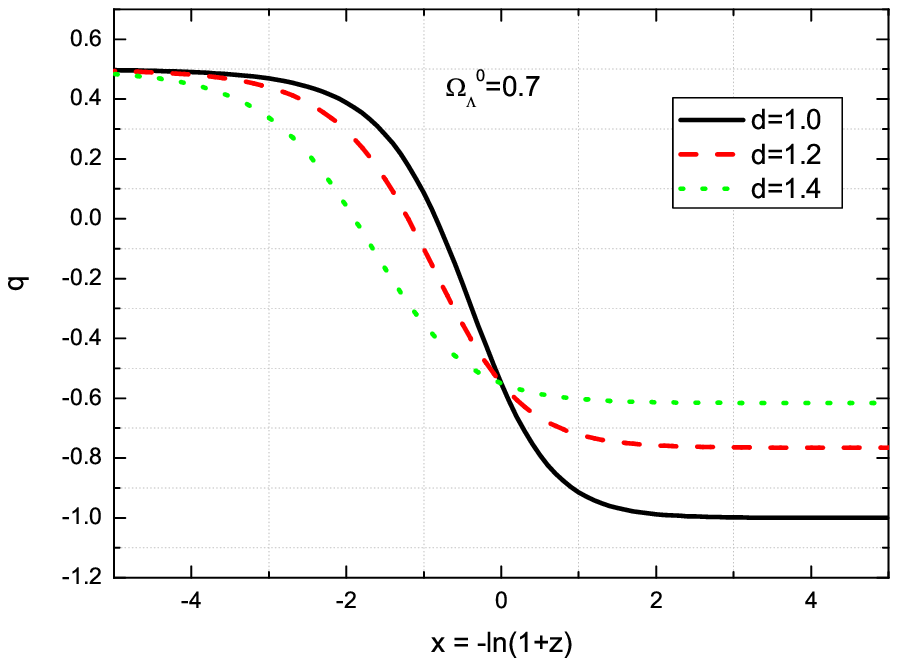}
\caption{Dependence of the deceleration parameter on the constant
$d$.} \label{fig:qx}
\end{center}
\begin{center}
\includegraphics[width=11cm]{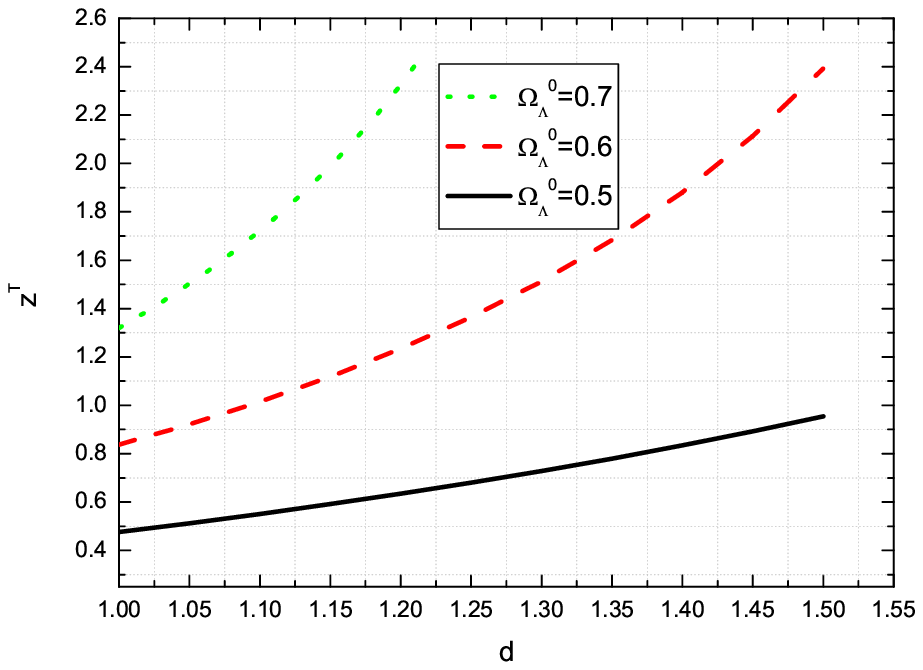}
\caption{Dependence of the deceleration parameter on the constant
$d$.} \label{fig:zTd}
\end{center}
\end{figure}
Fig.~\ref{fig:qx} shows the evolution of the deceleration
parameter $q$ where for definiteness the value of
$\Omega_\Lambda^0$ is set to be 0.7. The discussion above have
shown that in this case $d\in[ 1, 1.5 )$. From this figure, we can
easily see that $q_0=-0.5$ is independent of the parameter $d$ and
there is a transition from deceleration to accelerating expansion.
Fig.~\ref{fig:zTd} shows the relation between the redshift of the
turning point $z_T$ at which the deceleration expansion to
acceleration transition happened and the value of the parameter
$d$. It's obvious that the evaluation of present DE density
fraction imposes a practical constraint on the parameter $d$ which
is indicated with the rapidly ascending curve.


In conclusion, we have studied a kind of holographic DE model in
which the future event horizon is chosen to be the IR cutoff and
the equation of state is fixed to be $-1$. In this model, an
interaction between matter and dark energy naturally appears. We
find that the accelerating expansion as well as the transition
from deceleration to acceleration is well recovered. There exists
a stable tracker solution for the dimensionless parameter $d>1$.
So this model provides one possible phenomenological framework to
alleviate the cosmological coincidence problem with the
holographic motivation. We show that, by means of only one
cosmological parameter the DE density fraction, the constant $d$
obtains an observational upper bound. Specifically speaking, if
today's universe is DE dominated, there is a practical constraint
on the numerical factor $d$ which can not deviate much from $1$.
It's interesting to further examine this model with current
observational data and determine whether other strategies such as
a varying Newton's constant~\cite{Shapiro, Horvat, Bauer, HorvatG}
are necessary to extend our framework while making the
cosmological coincidence problem still ameliorated.

\section*{Acknowledgements}
We are grateful to Prof. R.G.Cai, M. Li, Dr. X.Zhang and D.F. Zeng
for useful discussions. This project was in part supported by
National Basic Research Program of China under Grant No.
2003CB716300 and by NNSFC under Grant No. 90403032.

\end{document}